\documentclass{aa}
\usepackage{graphicx}
\usepackage{natbib}
\bibpunct{(}{)}{;}{a}{}{,} 
\usepackage{txfonts}
\begin{document}
   \title{Low temperature Rosseland opacities with varied abundances of carbon and nitrogen}

   \author{M. T. Lederer
          \inst{1}
          \and
           B. Aringer
          \inst{1,2}
          }

   \offprints{M. T. Lederer}

   \institute{Department of Astronomy, University of Vienna,
              T\"urkenschanzstrasse 17, A-1180 Vienna, Austria\\
              \email{lederer@astro.univie.ac.at}
							\and
							Dipartimento di Astronomia, Universit\`a di Padova,
							Vicolo dell'Osservatorio 3, I-35122 Padova, Italy\\
							\email{aringer@astro.univie.ac.at}
             }

  \date{Received July 11, 2008; accepted October 31, 2008}

  \abstract
   {With certain assumptions, radiative energy transport can be modelled by the diffusion approximation. In this case, the Rosseland mean opacity coefficient characterises the interaction between radiation and matter. The opacity data are usually available in pre-tabulated form, and in the generation of the data one assumes a distinct heavy element mixture, which is usually a scaled solar one. Therefore, presently available data is unable to cover the full parameter range of some astrophysical problems, in which the chemical composition of the medium being considered varies.
   }
   {
We attempt to produce low temperature opacity data incorporating the effects of varied abundances of the elements carbon and nitrogen. For our temperature range of interest, molecules represent the dominant opacity source. Our dataset covers a wide metallicity range and is meant to provide important input data for stellar evolution models and other applications.
   }
   {
We conduct chemical equilibrium calculations to evaluate the partial pressures of neutral atoms, ions, and molecules. Based on a large dataset containing atomic line and continuum data and, most importantly, a plethora of molecular lines, we calculate Rosseland mean opacity coefficients not only for a number of different metallicities, but also for varied abundances of the isotopes \element[][12]{C} and \element[][14]{N} at each metallicity. The molecular data comprise the main opacity sources for either an oxygen-rich or carbon-rich chemistry. We tabulate the opacity coefficients as a function of temperature and, basically, density.
   }
   {
   Due to the special role of the CO molecule, within a certain chemistry regime an alteration to the carbon abundance causes considerable changes in the Rosseland opacity. The transition from a scaled solar (i.\,e. oxygen-rich) mixture to a carbon-rich regime results in opacities that can, at low temperatures, differ by orders of magnitude from to the initial situation. The reason is that the mean opacity in either case is due to different molecular absorbers. Variations in the abundance of nitrogen have less pronounced effects but, nevertheless, cannot be neglected.
   }
   {
In typical astrophysical applications, it is indispensable to take into account opacity variations due to chemistry changes. In this respect, the new data is superior to previous compilations, but is, however, still subject to uncertainties. 
   }

   \keywords{radiative transfer -- molecular data -- stars: evolution}

   \maketitle
%

\section{Introduction}\label{sec:introduction}
Radiative energy transport within a medium plays a role in a wealth of astrophysical scenarios. When the photon mean free path is short compared with the typical scale height of the medium, irrespective of the photon wavelength, this problem can be modelled by the diffusion approximation. 
By assuming local thermodynamic equilibrium (LTE), the definition of a harmonic mean opacity coefficient -- the so-called Rosseland mean \citep{1925ApJ....61..424R} -- emerges naturally from the evaluation of the integrated radiative flux. Rosseland stated that the derivation of the analytical expression for this "characteristic function of the medium regarding its power of transmitting radiation \textit{en bloc}" is not difficult, yet the actual calculation is. This is because one usually has to account in detail for a wide variety of absorbers. Typical applications of Rosseland mean opacities are stellar structure and evolution calculations, which span orders of magnitude in temperature and density and require the inclusion of numerous absorption and scattering processes. As an example, we consider a low mass star in the late phase of its evolution on the Asymptotic Giant Branch (AGB). Data for electron thermal conductivity, a basic energy transport process in the inert C-O core, is available from \citet{2007ApJ...661.1094C}. Tabulated high temperature opacity data (ranging from a few thousand to a few hundred million Kelvin) are provided by the OPAL collaboration \citep{1996ApJ...464..943I} and the Opacity Project \citep[OP, ][]{2005MNRAS.362L...1S}. Both groups provide their data publicly through web interfaces\footnote{OPAL: \texttt{http://www-phys.llnl.gov/Research/OPAL/}}$^,$\footnote{OP: \texttt{http://opacities.osc.edu}} where in general single element abundances can be varied. OPAL also allows to produce so-called Type II tables: starting from a given element mixture, the abundances of two elements (e.\,g. C and O) are enhanced by adding constant mass fractions in each case. Apart from H$_2$, however, no molecular contributions to the Rosseland mean are taken into account in both the OPAL and OP databases. Molecules become the dominant opacity source at temperatures lower than about $4000$-$5000\,\mathrm{K}$, which occur in the outer envelope and atmosphere of a star on one of the giant branches -- either the Red Giant Branch (RGB) or the AGB --, or in protoplanetary accretion disks. At even lower temperatures (below approximately $1500\,\mathrm{K}$), dust can be responsible for the bulk of opacity.

One of the first papers giving a detailed discussion about low temperature mean opacities (and also summarising earlier efforts on this topic) was \citet{1983ApJ...272..773A}. This work evolved further and resulted in the extensive database of \citet{1994ApJ...437..879A} and the updated version of \citet{2005ApJ...623..585F} -- henceforth AF94 and F05, respectively -- that has become a standard for low temperature opacities in the past few years. These data are based on scaled solar metal compositions (plus some tables for enhanced alpha element abundances). Returning to our AGB star example, the problem  is that the chemical composition in the envelope of such an object varies. Products of the ongoing nucleosynthesis in the stellar interior are brought to the surface by a series of mixing events. The entire mechanism is dubbed Third Dredge-up \citep[TDU, for a review see][]{1999ARA&A..37..239B}. The main burning products dredged up are freshly synthesised carbon and elements produced by the slow neutron capture process. Alterations to the element mixture result in significantly different opacity coefficients. The corresponding implications for the stellar evolution calculations were emphasised by \citet{1983ApJ...272..773A}. The authors provide some examples of how the Rosseland mean opacity changes when the number ratio of carbon to oxygen atoms (C/O) varies. The important role of the C/O ratio is due to the fact that of all the abundant molecules, carbon monoxide (CO) has the highest binding energy \citep{1934ApJ....79..317R}. The partial pressure of CO hardly varies with the C/O ratio in a plasma at constant temperature and pressure. In a chemical equilibrium situation, the less abundant species of carbon and oxygen are almost completely locked in the CO molecule. The remaining fraction of the more numerous atoms is free to form other molecules. The absolute number of free atoms (i.\,e. not bound in CO) of either oxygen or carbon primarily determines the magnitude and characteristics of the Rosseland mean opacity since CO contributes on average less than other molecular species (such as H$_2$O, TiO, C$_2$H$_2$, and HCN).

Although the above facts have been in principle known for a long time, they have not been accounted for in stellar evolution models, due to the lack of adequate opacity data. From the tabulated low temperature opacities that can be found in the literature \citep[other than AF94 and F05, e.\,g. ][partly monochromatic and focused on special objects such as brown dwarfs and extrasolar planets]{1993A&A...274..818N,2003A&A...410..611S,2006astro.ph..5666W,2007ApJS..168..140S,2008ApJS..174..504F}, only \citet{2000A&A...358..651H} considered a case in which $\mbox{C/O}=1.8$ in simulating winds of carbon-rich AGB stars. \citet{2007ApJ...666..403D} investigated single element abundance variations and the influence on low and high temperature opacities exemplarily. Subsequently, the sensitivity of stellar evolution models to these changes was analysed and both effective temperatures and stellar lifetimes depend clearly on opacity changes due to abundance variations.

A dramatic improvement in evaluating molecular opacities with a varying amount of carbon was achieved by \citet{2002A&A...387..507M} who illustrated how a correct treatment of opacity can affect stellar evolution tracks by using analytical AGB star models. The principle outcomes were that observations of carbon stars can be reproduced more accurately than before and the models appear to imply that the carbon star phase is shortened considerably with a consequent reduction in the stellar yields. The opacities were estimated by chemical equilibrium calculations and analytical fits for molecular contributions to the Rosseland mean, which were combined by a simple linear summation. This was the main problem of this approach. The non-linear nature of the Rosseland mean by definition renders the method of adding up opacity contributions a fragile approximation. Moreover, the derived molecular contributions are not unique because they cannot be determined in a non-ambiguous way. The Rosseland mean emphasises transparent spectral regions, and the gaps between absorption lines from a specific molecule can be filled by other species depending on the chemistry. Thus, this interplay crucially influences the total opacity. Finally, not all relevant molecular absorbers have been taken into account in the work described above.

From this overview, we conclude that the stellar evolution community require a complete, homogeneous database containing Rosseland mean opacity coefficients with varied abundances of carbon. The aim of this work is to fill this gap. We also included abundance variations in nitrogen, since in low and intermediate mass stars mechanisms might activate the CN cycle and trigger the reconversion of carbon isotopes into nitrogen. The corresponding mechanism acting in stars of mass lower than $4\,\mathrm{M_{\sun}}$ is dubbed Cool Bottom Process \citep{1995ApJ...447L..37W} which is a slow, deep circulation at the bottom of the convective envelope. Its origin is still unknown, although a number of driving mechanisms have been proposed (e.\,g. rotationally induced instabilities, magnetically induced circulation, gravity waves or thermohaline mixing -- we refer to \citealp{2007ApJ...671..802B} for an overview). In more massive AGB stars ($M>4\,\mathrm{M_{\sun}}$), the analogous process is known as Hot Bottom Burning \citep{1973ApJ...185..209I}.

A preliminary version of the opacity data presented in the following was applied successfully by \citet{2007ApJ...667..489C,2008AIPC.1001....3C} and \citet{2008arXiv0805.3242L}. \citet{2007ApJ...667..489C} considered a low metallicity ($Z=1\times10^{-4}$) stellar model with $M=2\,\mathrm{M_{\sun}}$. By using the new opacity coefficients, the model was able to reproduce observational data more accurately than earlier models in terms of physical properties (such as for instance the effective temperature) and the abundance pattern of the heavy elements. Information that is complementary to the data considered here can be found in \citet{2007AIPC.1001...11L}.

The paper is structured as follows. In Sect.~\ref{sec:toolandmethod}, we describe the tools used to generate our opacity tables. Section~\ref{sec:datasources} summarises all data adopted in our calculations, i.\,e. abundances as well as atomic and molecular opacity data. We describe and justify the design of our tables in Sect.~\ref{sec:tabledesign}. We discuss the results in detail in Sect.~\ref{sec:results}, before concluding and providing a perspective on future work in Sect.~\ref{sec:conclusions}.

\section{Tool and method}\label{sec:toolandmethod}
To generate the data presented in this work, we used the COMA code developed by \citet{Aringer2000}. For a description of improvements since the initial version, we refer to a forthcoming paper of Aringer et al. (in preparation). Assuming local thermodynamic equilibrium, the program solves for the ionisation and chemical equilibrium (using the method of \citealp{1973A&A....23..411T}) at a given temperature and density (or pressure) combination for a set of atomic abundances. From the resulting partial pressures for the neutral atoms, ions, and molecules, the continuous and line opacity is calculated at the desired wavelengths using the data listed in Sect.~\ref{sec:datasources}. The main purpose of the COMA code was originally to provide monochromatic absorption coefficients for dynamical model atmospheres of cool giants \citep{2003A&A...399..589H}. However, it has been used in a wide range of applications, for example in the process of calculating low and high resolution spectra \citep[e.\,g. ][respectively]{2002A&A...395..915A,2008arXiv0805.3242L} and line profile variations \citep{2005A&A...437..273N}.

Once the monochromatic opacities $\kappa_\nu(T,\rho)\equiv\kappa(\nu,T,\rho)$ are known ($T$ stands for the temperature and $\rho$ for the density), the calculation of a mean opacity coefficient such as the Rosseland mean is straightforward. One has to perform a weighted integration of $\kappa_\nu$
over the relevant frequency range. For the Rosseland mean $\kappa_\mathrm{R}=\kappa_\mathrm{R}(T,\rho)$, the relation is given by
\begin{equation}
\label{eq:rosselandmean}
\frac{1}{\kappa_\mathrm{R}}=\int _0 ^{\infty} \frac{1}{\kappa _\nu} \frac{\partial B_\nu(T)}{\partial T}d\nu\ \Big/ \int _0 ^{\infty} \frac{\partial B_\nu(T)}{\partial T} d\nu\mbox{,}
\end{equation}
where the weighting function is the partial derivative of the Planck function with respect to the temperature. The main subject of this paper is to study not only the dependence of $\kappa_\mathrm{R}$ on the thermodynamic quantities $T$ and $\rho$, but in addition the chemical composition.

In practice, the integration over $\nu$ in Eq.~\ref{eq:rosselandmean} must be performed at a predefined discrete frequency (or wavelength) grid. We use a grid that is based on the one described by \citet{1992A&A...261..263J}, but we extend the wavelength range to have boundaries at $200{,}000\,\mathrm{cm}^{-1}$ ($500\,\AA=0.05\,\mu\mathrm{m}$) and $200\,\mathrm{cm}^{-1}$ ($50\,\mu\mathrm{m}$). This results in a total number of 5645 wavelength points at which we calculate the opacity. \citet{1998A&A...337..477H} proposed to use a number of opacity sampling points for the accurate modelling of atmospheres of late-type stars that is roughly four times larger than the value used here. The same is true for the number of wavelength points in F05, who use more than $24{,}000$ points. However, the error introduced when using a lower resolution is generally small compared with other uncertainty sources (see Sect.~\ref{sec:uncertainties}), and a smaller number of grid points has the advantage of a reduced computing time. We note that the opacity sampling technique remains -- regardless of the precise number of points -- a statistical method. To arrive at a realistic and complete description of the spectral energy distribution, one would need a far higher resolution ($R\simeq200{,}000$). In any case, the grid is sufficiently dense for a rectangle rule to be sufficient in carrying out the wavelength integration. This can be justified by comparing the numerically obtained value of the normalisation factor on the right-hand side of Eq.~\ref{eq:rosselandmean} with its analytical value. The formal integration limits in the definition of the Rosseland mean have to be replaced by cut-off wavelengths. These are determined by the weighting function $\partial B_\nu(T)/\partial T$ that constrains the relevant wavelength range. Like the Planck function itself, the maximum of its derivative shifts to higher wavelengths with decreasing temperature and vice versa. At the upper wavelength limit adopted here ($50\,\mu\mathrm{m}$) and for the lowest temperature at which we generate data (about $1600\,\mathrm{K}$), the weighting function has decreased by more than $1/100$ relative to its maximum value. Accordingly, at the high temperature edge the weighting function at the same wavelength has dropped to almost $1/10{,}000$ of its maximum value. Since we do not include grain opacity in our calculations that would require going to even higher wavelengths (as in F05), we definitely cover the relevant spectral range for the calculation of $\kappa_\mathrm{R}$ within the adopted parameter range. The decline in the weighting function towards lower wavelengths (or higher frequencies) is far steeper, so that the above argument is also fulfilled at the low-wavelength cut-off.

\section{Data sources}\label{sec:datasources}
In the following, we briefly summarise the sources of the data entering the opacity calculations. The basic ingredient in this procedure is the relative amount of elements contained in the mixture for which we would like to know the opacity. In this work, we chose to use the set of recommended values for solar element abundances compiled by \citet{2003ApJ...591.1220L}, which imply a solar C/O ratio of $0.501$. The values for the elements C, N, and O are close to the values given by \citet{2007SSRv..130..105G}; using their abundances results in (C/O)$_\odot=0.537$. The authors derive these values from a 3D hydrodynamic model of the solar atmosphere, a technique that caused a downward revision of the solar CNO abundances in recent years. However, these values are still disputed. For instance, \citet{2008ApJ...682L..61C} used spectro-polarimetric methods to argue for an oxygen abundance that is higher than claimed by \citet{2007SSRv..130..105G} and closer to previously accepted values (e.\,g. \citealp{1998SSRv...85..161G} with (C/O)$_\odot=0.490$) that agree with values derived from helioseismology (e.\,g. \citealp{2008PhR...457..217B}).

However, the data presented depend to first order only on the relative amount of carbon and oxygen, which does not differ significantly from that for the various abundance sets mentioned above. Therefore and due to the fact that C/O is a variable quantity, the current tables can serve as an approximation for applications that use abundances other than \citet{2003ApJ...591.1220L}, until we generate further data.

\subsection{Continuous opacity}

\begin{table}
\caption{Continuous opacity sources}
\centering
\label{table:continuum}
\begin{tabular}{l l}
\hline\hline
Ion and process & Reference\\
\hline
\ion{H}{i} b-f and f-f & \citet{1961ApJS....6..167K}\\
H$^-$ b-f              & \citet{1966MNRAS.132..255D}\\
H$^-$ f-f              & \citet{1966MNRAS.132..267D}\\
H+H (Quasihydrogen)    & \citet{1968ApJ...153..987D}\\
H$_2^+$ f-f            & \citet{1965ApJS....9..321M}\\
H$_2^-$ f-f            & \citet{1964ApJ...139..192S}\\
\ion{C}{i}, \ion{Mg}{i}, \ion{Al}{i}, \ion{Si}{i}, \ion{He}{i} f-f & \citet{1970MmRAS..73....1P}\\
He$^-$ f-f                          & \citet{1965ApJ...141..811S}\\
continuous $e^-$ scattering         & \citet{1978stat.book.....M}\\
Rayleigh scattering from \ion{H}{i} & Dalgarno, quoted by \citet{1964SAOSR.167...17G}\\
Rayleigh scattering from H$_2$      & \citet{1962ApJ...136..690D}\\
\hline
\end{tabular}
\end{table}

The routines for the calculation of the continuum opacity in COMA are based on an earlier version of the MARCS code \citep{1992A&A...261..263J}. The latest MARCS release was described by \citet{2008A&A...486..951G}. We adopt the format of their Table 1 to ease comparison with their updated set of continuous opacity sources. The data that we use (and list in Table~\ref{table:continuum}) is not as extensive. However, the most relevant sources are included, and in the low temperature region of the presented tables in particular the molecular contribution to $\kappa_\mathrm{R}$ dominates over the continuum by several orders of magnitude.

\subsection{Atomic lines}
Atomic line data are taken from the VALD database \citep{2000BaltA...9..590K}\footnote{VALD: \texttt{http://ams.astro.univie.ac.at/\~{}vald/}}, where we use updated version 2 data (from January 2008) here. For the  atomic lines, we adopt full Voigt profiles derived from the damping constants listed in VALD. The only exception is hydrogen for which we use an interpolation in tabulated line profiles from \citet{1995yCat.6082....0S}. The atomic partition functions are taken from the work of \citet{1981ApJS...45..621I}, although the data for boron has been modified \citep{Gorfer2005}. The number of atomic lines included is $16{,}059{,}201$. Split into their respective ionisation stages the numbers are: \ion{}{I} $4{,}028{,}995$, \ion{}{II} $5{,}347{,}990$, and \ion{}{III} $6{,}682{,}216$. For the ionisation energies, we refer to \citet{Stift2000} since we use the same reference data quoted there in the current work.

\subsection{Molecular data}

\begin{table}
\begin{minipage}[t]{\columnwidth}
\caption{Molecular line data}
\label{table:molecules}
\centering
\renewcommand{\footnoterule}{}                          
\begin{tabular}{l c r c}
\hline\hline
Molecule & Thermodynamic         & Number of lines & Line \\
         & data                  
           \footnote{
		(1) \citealt{1984ApJS...56..193S},
                (2) \citealt{1985A&A...148...93R},
		(3) \citealt{VidlerTennyson2000},
                (4) \citealt{2002JChPh.11711239B},
		(5) \citealt{1988A&AS...74..145I},
                (6) \citealt{2003ApJ...594..651D}
	   }
				&                 & data
						     \footnote{
							(7) \citealt{1994ApJS...91..483G},
							(8) \citealt{Jorgensen1997}, 
							(9) \citealt{1974A&A....31..265Q},
						       (10) \citealt{LanghoffBauschlicher1993},
						       (11) \citealt{1998cpmg.conf..321S},
						       (12) \citealt{2006MNRAS.368.1087B},
						       (13) \citealt{2006MNRAS.367..400H}, 
						       (14) \citealt{Schwenke1997},
						       (15) \citealt{1998A&A...330.1109A},
						       (16) \citealt{1995SPIE.2471..105R},
						       (17) \citealt{2005JQSRT..96..139R},
						       (18) \citealt{Tipping2007},
						       (19) \citealt{Plez2007},
						       (20) \citealt{Littleton1987},
						       (21) \citealt{BauschlicherRam2001},
						       (22) \citealt{1989ApJ...343..554J}
						     }
						          \\

\hline
CO         & 1 &    	131\,859 &  7 \\
CH         & 2 &    	229\,134 &  8 \\
C$_2$      & 1 &    	360\,887 &  9 \\
SiO        & 2 &     	 85\,788 & 10 \\
CN         & 1 &     2\,533\,040 &  8 \\
TiO        & 1 &    22\,724\,670 & 11 \\
H$_2$O     & 3 &    27\,693\,367 & 12 \\
HCN/HNC    & 4 &    33\,454\,021 & 13 \\
OH         & 1 &         36\,601 & 14 \\
VO         & 1 &     3\,171\,552 & 15 \\
CO$_2$     & 2 &     1\,032\,266 & 16 \\
SO$_2$     & 5 &         29\,559 & 17 \\
HF         & 1 &             462 & 18 \\ 
HCl        & 1 &             447 & 17 \\ 
ZrO        & 1 &    16\,391\,195 & 19 \\
YO         & 1 &             975 & 20 \\
FeH        & 6 &        116\,300 &  6 \\
CrH        & 1 &         13\,824 & 21 \\
\hline
C$_2$H$_2$ & - & opacity sampling & 22 \\
C$_3$      & - & opacity sampling & 22 \\
\hline
\end{tabular}
\end{minipage}
\end{table}

In the calculation of low temperature opacities, molecules play a critical role. We list the data that we used to calculate the tables containing Rosseland mean opacity coefficients in Table~\ref{table:molecules}. References to the thermodynamic data (i.\,e. the partition function) and the line data for each molecule (displayed in the first column) are as indexed in the columns two and four, respectively. The number of lines entering the calculation is given in the third column (the original lists contained more lines). For some molecules, there is more than one line list available. The line lists that we use were selected during other projects (cited at the beginning of Sect.~\ref{sec:toolandmethod}) proven to deliver viable results. In the case of the OH line list, the measured data from the HITRAN database contains more lines than that from \citet{Schwenke1997}. We performed some test calculations after completing the database with OH data from HITRAN. The change in our results was, however, hardly perceptible. As long as the overall opacity distribution is reproduced reasonably, the line positions do not have to be precisely correct in calculating $\kappa_\mathrm{R}$ (as opposed to applications relying on high-resolution spectral synthesis).

\begin{table*}
\caption{Metallicities and enhancement factors}
\label{table:metallicities}
\centering
\begin{tabular}{l | c c c c c c c c c | c c c}
\hline\hline
\multicolumn{1}{l}{Metallicity (Z)} & \multicolumn{9}{c}{\element[][12]{C} enhancement factors} & \multicolumn{3}{c}{\element[][14]{N}} enhancement factors\\
\hline
0.04& 1.0 & 1.2 & 1.5 & 1.8 & 2.0 & 2.2 & 3.0	& & & 1.0	& 1.2 &	1.5 \\                            
0.02& 1.0 & & 1.5 & 1.8 & 2.0 & 2.2 & 3.0 & 5.0	& & 1.0	& 1.5 &	2.0 \\
0.01& 1.0 & & 1.5 & 1.8 & 2.0 & 2.2 & 5.0 & 10.0	& & 1.0 & 1.5 &	3.0		\\
0.008& 1.0 & & & 1.8 & 2.0 & 2.2 & 5.0 & 10.0	& 20.0 & 1.0 &	2.0 & 4.0		\\
0.006& 1.0 & & & 1.8 & 2.0 & 2.2 & 5.0 & 10.0	& 30.0 & 1.0 & 2.5 & 5.0		\\
0.005& 1.0 & & & 1.8 & 2.0 & 2.2 & 5.0 & 10.0	& 30.0 & 1.0 & 2.5 & 5.0		\\
0.004& 1.0 & & & 1.8 & 2.0 & 2.2 & 5.0 & 10.0	& 30.0 & 1.0 & 2.5 & 5.0		\\
0.003& 1.0 & & & 1.8 & 2.0 & 2.2 & 6.0 & 20.0	& 70.0 & 1.0 & 3.0 & 8.0		\\
0.002& 1.0 & & & 1.8 & 2.0 & 2.2 & 6.0 & 20.0	& 70.0 & 1.0 & 3.0 & 8.0		\\
0.001& 1.0 & & & 1.8 & 2.0 & 2.2 & 8.0 & 35.0	& 150.0 & 1.0 & 4.0 & 16.0		\\
0.0003& 1.0 & & & 1.8 & 2.0 & 2.2 & 12.0 & 75.0	& 500.0 & 1.0 & 7.0 & 50.0		\\
0.0001& 1.0 & & & 1.8 & 2.0 & 2.2 & 18.0 & 150.0	& 1500.0 & 1.0 & 12.0 & 150.0		\\
0.00003& 1.0 & & & 1.8 & 2.0 & 2.2 & 27.0 & 350.0	& 5000.0 & 1.0 & 22.0 & 500.0		\\
0.00001& 1.0 & & & 1.8 & 2.0 & 2.2 & 40.0 & 750.0	& 15000.0 & 1.0 & 40.0 & 1500.0 	\\
\hline
\end{tabular}
\end{table*}

For the molecules C$_2$ and CN, we introduced some modifications to the line data. \citet{Jorgensen1997} suggested a scaling of the $gf$ values for C$_2$ in a certain wavelength region (details are given in \citealp{2001A&A...371.1065L}). We followed this suggestion but also investigated the effect on $\kappa_\mathrm{R}$ of not applying this scaling (see Sect.~\ref{sec:uncertainties}). The modifications to the CN line list are not crucial to the calculation of the Rosseland mean. We corrected the line positions of approximately $18{,}000$ lines using measurements listed in the catalogue of \citet{2005cns..book.....D}. The data for the molecules C$_3$ and C$_2$H$_2$ from \citet{Jorgensen1997} are only available in the form of an opacity sampling. For the generation of this data, a microturbulent velocity of $2.5\,\mathrm{km\,s^{-1}}$ was adopted. As for the calculation of the molecular line opacity, we assume Doppler profiles in each case, since there is little information about the damping constants of molecular transitions. Additionally, for the species dominating the overall opacity and in regions close to bandheads, even the wings of the strongest lines will contribute less to the opacity than the Doppler cores of the numerous overlapping neighbour lines.

The chemical equilibrium constants used in calculating of the molecule partial pressures are those from \citet{1973A&A....23..411T} with updates described in \citet{1996A&A...315..194H}. In the case of C$_3$, we utilised data published by \citet{1981ApJS...45..621I}. All molecules (a total number of 314 species) enter into the equation of state.

\section{Table design}\label{sec:tabledesign}
We tabulate the logarithm of the Rosseland mean opacity $\log \kappa_\mathrm{R}\,\mathrm{[cm^2\,g^{-1}]}$ as a function of $\log T \mathrm{[K]}$, the logarithm of the gas temperature, and $\log R\,\mathrm{[g\,cm^{-3}\,K^{-3}\,10^{18}]}$, where $R\equiv\rho/T_6^3$. $\rho$ and $T_6$ represents the density in units of $\mathrm{g\,cm^{-3}}$ and the temperature in millions of Kelvin, respectively. The ranges covered are $3.2 \leq \log T\,\mathrm{[K]} \leq 4.05$ with a step size of $0.05$, and $-7.0 \leq \log R \leq 1.0$ with a step size of $0.5$. The low temperature cut-off was set to be the temperature at which grains may become the major opacity source (cf. e.\,g. F05). However, typical AGB stellar evolution models do not attain such low temperatures. We would like to emphasise that dust formation is usually not an equilibrium process \citep{1988A&A...206..153G}, and thus an a priori tabulation of grain opacities can be affected by large uncertainties. At the highest value of $\log T$, the contribution of molecules to the mean opacity vanishes and a smooth transition to high temperature opacity data is possible (see Sect. \ref{sec:comparison}). Data are available for 14 different values of $Z$ (cf. Table~\ref{table:metallicities}) that depicts the total mass fraction of all elements heavier than helium. The metallicity spans a range from $Z=1\times10^{-5}$ to the Magellanic clouds metallicity range \citep{2000glg..book.....V,2000PASP..112..529V}, where the grid is a bit denser in terms of $Z$, and the solar metallicity to super-solar metallicity ($Z=0.04$). We calculated tables for three different mass fractions of hydrogen ($X\in\left\{0.5,0.7,0.8\right\}$). Thus, the data cover the supposedly prior field of application, namely the outer layers of an AGB star, where such values of $X$ occur. Initially, we calculated a master table for each metallicity, in which the metal composition was linearly scaled from the abundances given by \citet{2003ApJ...591.1220L} to arrive at the respective $Z$ value. From these abundances, we then enhanced the mass fractions of \element[][12]{C} and \element[][14]{N} in steps that depended on the initial metallicity (see Table~\ref{table:metallicities}). Since this produced an increase in the overall metallicity $Z$, we followed the OPAL approach and reduced the mass fraction of \element[][4]{He} to fulfil $X+Y+Z=1$.

The number of enhancement factors is constrained by a trade-off between numerical accuracy and computational costs in the generation and application of the tables. Due to the special role of carbon in conjunction with oxygen (see Introduction), the number of enhancement factors is higher than for nitrogen.\footnote{Here we deviate from the original design used in \citet{2007ApJ...667..489C} and \citet{2007AIPC.1001...11L}, where we adopted 5 enhancement factors for both carbon and nitrogen .} We calculated opacities for 7 different mass fractions of \element[][12]{C}. The starting point was the mass fraction that results from scaling all element abundances to the metallicity under consideration. All other carbon mass fractions resulted from multiplying this mass fraction by factors chosen as follows. For each metallicity, we use the factors 1.8, 2.0, and 2.2. The C/O ratio emerging from the adopted scaled solar abundances was about 0.5. Thus, multiplying the initial $X(\mathrm{\element[][12]{C}})$ by 2.0, we inferred that $\mbox{C/O}\simeq1$, where the molecular opacity in general reaches a minimum at low temperatures. As one can see for example in Fig.~\ref{fig:kappa-c-logT-logRm1p5}, the molecular absorption increased sharply on both sides of this minimum, while, towards much higher and lower C/O ratios, there was some type of saturation. To resolve this sharp turnaround, we included the factors 1.8 and 2.2 (corresponding to $\mbox{C/O}\simeq0.9$ and $1.1$, respectively). In stellar evolution models of low metallicity AGB stars this is probably of minor importance because the transition to a carbon star is usually rapid and occurs within a few dredge-up episodes \citep[e.\,g. ][]{2007A&A...469..239M}. The respective highest enhancement factor is related to the initial metallicity. From the work of \citet{2003PASA...20..314A} and references therein and the first application of our data by \citet{2007ApJ...667..489C,2008AIPC.1001....3C}, we derived information about the final carbon abundances reached in AGB stars. These determined the maximum enhancement factors in our tables for the metallicities under consideration (for instance $Z=1\times10^{-2},1\times10^{-3},1\times10^{-4}$). At all other metallicities, we derived the highest factors using a roughly linear relation between $\log Z$ and $\log f_\mathrm{C,max}$ (where we denote the enhancement factor as $f$). The remaining two factors for carbon are distributed almost equispaced on a logarithmic scale between $f=2$ and $f_\mathrm{C,max}$ at the lower metallicities. For high $Z$ with low final enhancements, we shifted factors from the carbon-rich regime to the region in which $\mbox{C/O}<1$. For \element[][14]{N}, we introduced two additional factors beyond the initial abundance. The expected final overabundance of nitrogen is much lower than for carbon, and we set $f_\mathrm{C,max}/f_\mathrm{N,max}=10$ for the lowest values of $Z$, decreasing this value for increasing $Z$. The intermediate enhancement factor for nitrogen was set to bisect the logarithmic interval approximately between the two other factors. An overview of all enhancement factors at each metallicity is given in Table~\ref{table:metallicities}.

We have not yet considered varied alpha element abundances, although one expects an increased value of [O/Fe] at low metallicities (e.\,g. \citealp{2000A&A...364L..19H} and references therein). The reason is the sharp increase in the number of tables, if one retains the previously outlined data structure. In its current version, the database contains $3\times7\times3=63$ tables per metallicity. Varying the abundance of oxygen also influences the C/O ratio, which is the decisive quantity for the molecular opacity at low temperatures. This in turn requires alterations to the enhancement factors of carbon, since one wishes to retain at least one point where $\mbox{C/O}=1$, even for an enhanced oxygen abundance. Establishing a scheme with a minimal number of enhancement factors for three elements (or element groups) is therefore not straightforward if one is attempting to retain as much information as possible with respect to the role of the C/O ratio. In place of enhancement factors, one could add constant amounts of one element (group) in terms of a mass fraction, as completed in the OPAL Type II tables. Only the application of the data in its current form will provide us with information about the feasibility of our approach and whether the data should be arranged in a different way. Future work will be dedicated to these questions.

\section{Results and discussion}\label{sec:results}

\begin{figure*}
   \centering
   \resizebox{\textwidth}{!}{\includegraphics{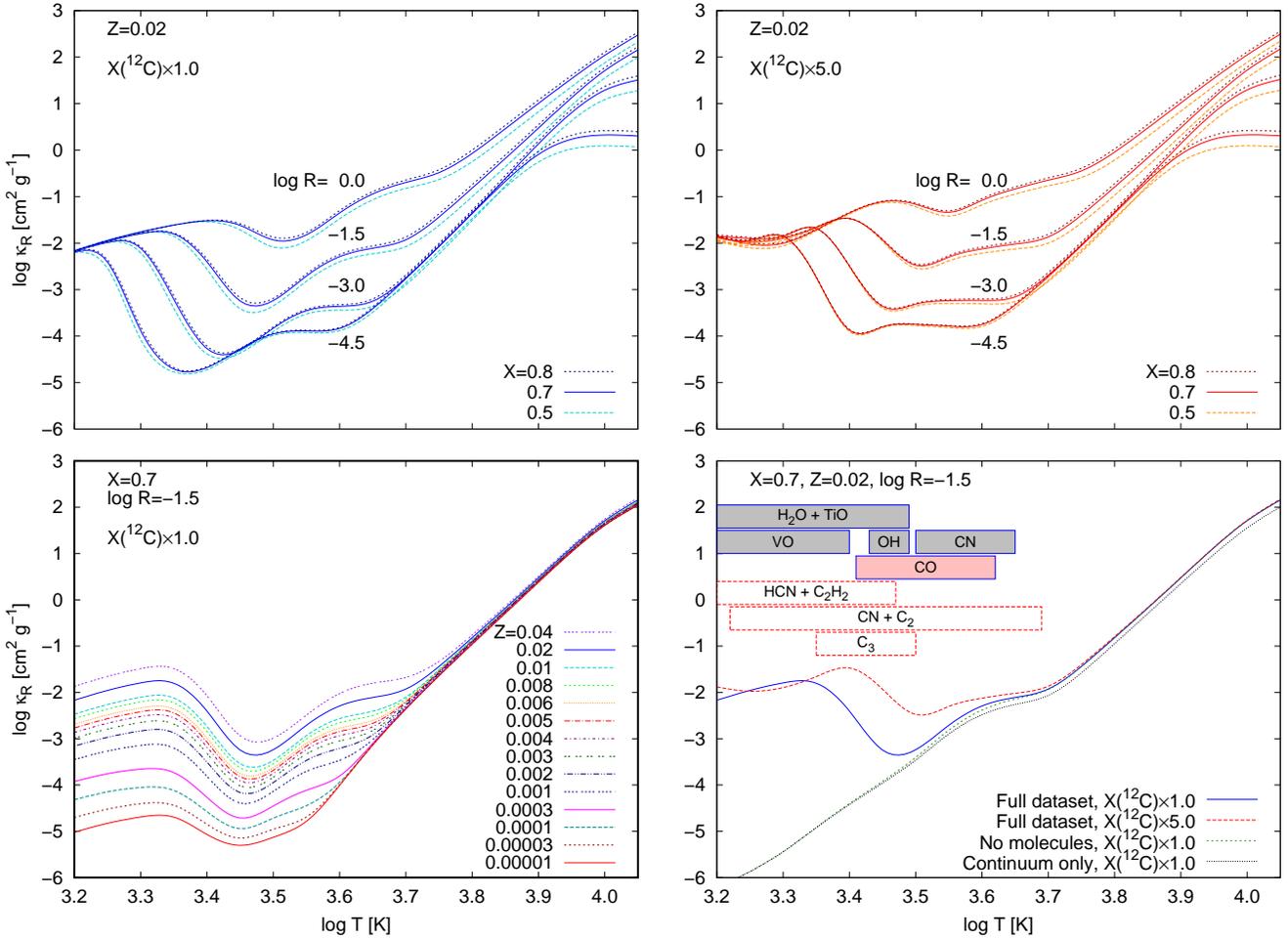}}
   \caption{Contents of the opacity database in some showcases. Top left panel: Rosseland mean opacities at constant $\log R$ for $Z=0.02$ as a function of $\log T$ for different values of $X$. The qualitative behaviour is fairly independent of $X$ and $Z$ and varies smoothly with $\log R$ at fixed abundances of carbon and nitrogen (the latter is not enhanced in any of the panels shown in this figure). Bottom left panel: The molecular opacity decreases when lowering the metallicity, but the structure with a bump at low temperatures due to the molecular contribution to the opacity is conserved. Top right panel: For the case where carbon is enhanced the molecules also produce high opacities at low $\log T$, although the shape of the curve differs noticeably from the standard case. Bottom right panel: The reason for the different structure in the opacities is that different molecules contribute to the opacity in the carbon-rich case (white boxes), in contrast to the oxygen-rich case (dark-grey boxes). The CO molecule contributes in both cases as well as CN, although at different orders of magnitude in each case. The extension of the boxes provide an indication of the temperature where the sources contribute but does not contain information about the order of magnitude of the contribution. The total contribution of molecules and atoms is assessed by leaving out these opacity sources in the calculations. All curves have been smoothed using cubic splines. See text for details.}
   \label{fig:coma-x-z-showcase-4up}
\end{figure*}

The calculation of Rosseland mean opacities for scaled solar metal mixtures has been discussed extensively in many papers (see Introduction for citations). Since we find good agreement with data from other groups (see Sect.~\ref{sec:comparison}), we only briefly restate the main points of this procedure. In Fig.~\ref{fig:coma-x-z-showcase-4up}, we provide an overview of the contents of the database using examples. We refer to individual panels in the following paragraphs. Generally speaking, molecules cause the mean opacity to vary dramatically as a function of temperature in a similar way for each hydrogen mass fraction $X$ and metallicity $Z$ (left panels of Fig.~\ref{fig:coma-x-z-showcase-4up}) that we include in our database. Beyond a temperature corresponding to $\log T=3.6$ to $3.7$ (depending on $Z$ and $\log R$), the contribution of molecules to $\kappa_\mathrm{R}$ vanishes and only continuous sources and atomic lines block the radiation field. 

AF94 provided an in-depth discussion about which type of opacity is dominant in different regions of the parameter space. They presented a detailed treatment of the main molecular opacity sources, which were in this case water (H$_2$O) -- accounting for the large bump at low temperatures -- and titanium oxide (TiO). They also provided information about the monochromatic absorption coefficients of these molecules. Beside H$_2$O and TiO, there is a number of other molecules that deliver non-negligible contributions to the Rosseland opacity in the oxygen-rich case \citep[cf. ][]{2007AIPC.1001...11L}. At higher temperatures, CN and CO contribute to $\kappa_\mathrm{R}$, and VO, OH, and SiO (ordered by decreasing importance) should also be taken into account. Calculations based on these 7 molecules result in opacity coefficients that are accurate to about 10 per cent compared with the full dataset when the metal mixture is oxygen-rich. The further inclusion of CrH and YO reduces this error to below 3 per cent on average. The temperature ranges in which different molecules contribute to $\kappa_\mathrm{R}$ indicated in the bottom right panel of Fig.~\ref{fig:coma-x-z-showcase-4up} are estimated by omitting the respective molecules in the calculation of $\kappa_\mathrm{R}$ and by checking where the change in this quantity exceeds 5 per cent with respect to the complete dataset. However, the limits derived from this criterion vary with the value of $\log R$ under consideration. We use $\log R=-1.5$, which is typical of the envelopes of AGB stars (Sergio Cristallo, private communication).

\begin{figure*}
   \centering
   \resizebox{\textwidth}{!}{\includegraphics{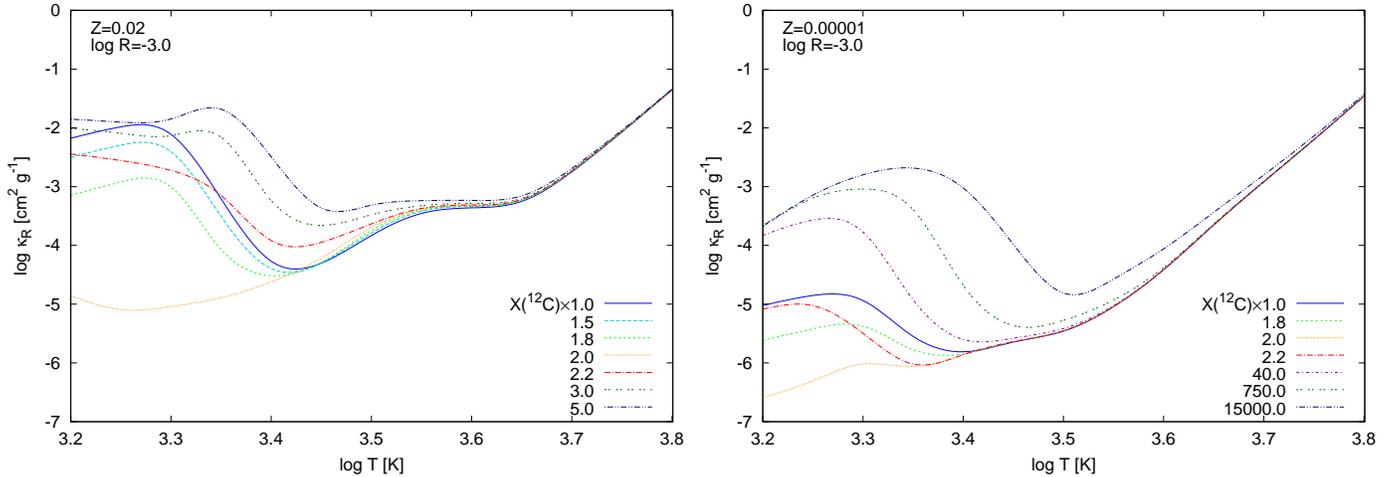}}
   \caption{Rosseland opacity changing with the carbon content in the metal mixture as a function of temperature. We show curves for two metallicities, $Z=0.02$ and the lowest metallicity in this database $Z=0.00001$, at a value of $\log R=-3.0$. The full line represents the solar scaled metal mixture with $\mbox{C/O}\simeq0.5$. An increase in the carbon mass fraction first causes a drop in $\kappa_\mathrm{R}$ at low temperatures as C/O approaches 1 ($X(\mathrm{\element[][12]{C}})\times2.0$, dotted line), because more oxygen atoms get bound in CO and less other molecular opacity carriers can be formed. When C/O rises beyond 1 this trend is reversed and the mean opacity increases again due to the formation of carbon-bearing molecules. At higher temperatures the opacity is growing monotonically due to CO, CN and atomic carbon. See also Fig.~\ref{fig:kappa-c-logT-logRm1p5}. All curves have been smoothed using cubic splines.
}
   \label{fig:coma-c-enhancment-Zhilo-logRm3p0}
\end{figure*}

\begin{figure*}[!ht]
   \centering
   \resizebox{\textwidth}{!}{\includegraphics{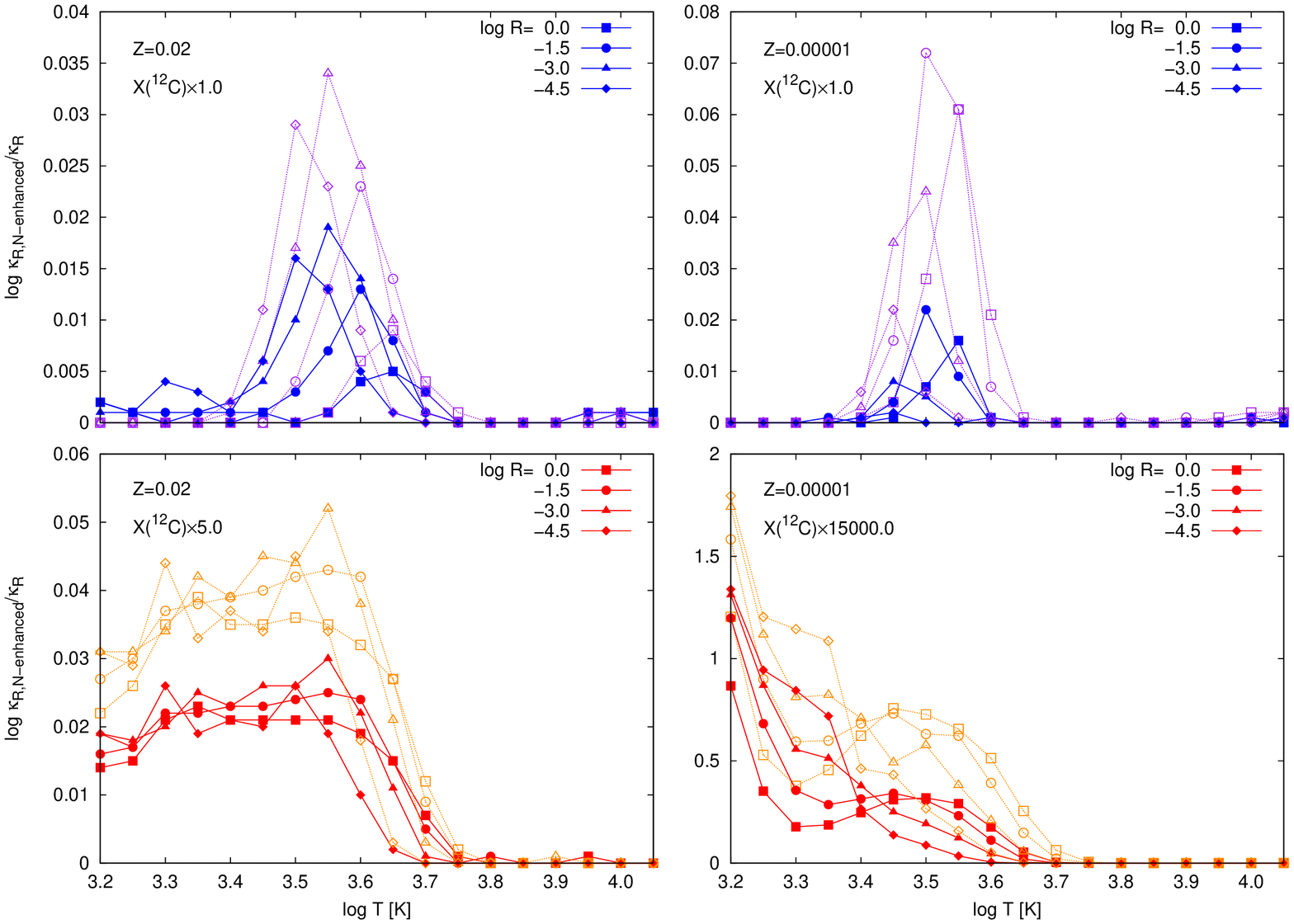}}
   \caption{Effects of an increased nitrogen abundance on the Rosseland mean opacity at various $\log R$ values relative to the case without any nitrogen enhancement. The filled and empty symbols refer to intermediate and maximum enhancement factors of \element[][14]{N}, respectively. For $Z=0.02$ (left panels), these are $1.5$ and $2$, whereas for $Z=0.00001$ (right panels), we have $40$ and $1500$. On the top, we show the cases for the oxygen-rich case (default carbon abundance). Here the increase in opacity is due to the CN molecule. The bottom panels refer to maximum enhanced \element[][12]{C} (i.\,e. the carbon-rich case), where HCN also contributes to the mean opacity at lower temperatures. Note the different scales on the y-axis.
}
   \label{fig:coma-nitrogen-relative-4}
\end{figure*}

An increase in the carbon content of the metal mixture while the amount of oxygen remains at its original value (i.\,e. an increase in the C/O ratio) has distinct effects on the Rosseland mean opacity. We refer to the Introduction and the previous section for a description of the respective mechanism. In the transition from the oxygen-rich regime (Fig.~\ref{fig:coma-x-z-showcase-4up}, top left) to the carbon-rich regime (Fig.~\ref{fig:coma-x-z-showcase-4up}, top right), one can distinguish between two cases. At lower temperatures (below $\log T=3.4$; we refer in the following to the case shown in Fig.~\ref{fig:coma-c-enhancment-Zhilo-logRm3p0}, i.\,e. $\log R=-3.0$), the opacity first decreases, due to the above described property of CO, which causes the following mechanism. As C/O increases from its initial value of about $0.5$ and approaches 1, more oxygen becomes bound in CO and fewer oxygen atoms are free to form molecules with a large overall absorption such as H$_2$O. Close to $\mbox{C/O}=1$ (not necessarily at an exact equal amount of C and O), the opacity reaches a minimum as the partial pressures of oxygen-bearing molecules drop substantially, while the abundances of carbon-bearing molecules only begin to rise to significant levels. As the amount of carbon continues to increase, thus incrementing C/O beyond 1, the opacity increases due to the formation of polyatomic carbon-bearing molecules such as C$_2$H$_2$ or HCN. These polyatomic molecules are obviously most relevant at lower temperatures. In addition, C$_3$ and C$_2$ produce a bump in the opacity at high carbon enrichments.

The situation at higher temperatures (up to $\log T=3.7$) is, however, different. Besides CO, the contribution of which remains almost constant, only CN and C$_2$ are relevant opacity sources (cf. Fig.~\ref{fig:coma-x-z-showcase-4up}, bottom right panel), while other molecules that are important to the Rosseland opacity are dissociated at these temperatures. The partial pressures of these molecules rise monotonically with the carbon abundance and thus the opacity at high $\log T$ increases in the same manner. At intermediate temperature (around $\log T=3.4$) where many different opacity sources contribute, the aforementioned mechanisms compete and the behaviour of $\kappa_\mathrm{R}$ is a more complex function of C/O.

\begin{figure}
   \centering
   \resizebox{\columnwidth}{!}{\includegraphics{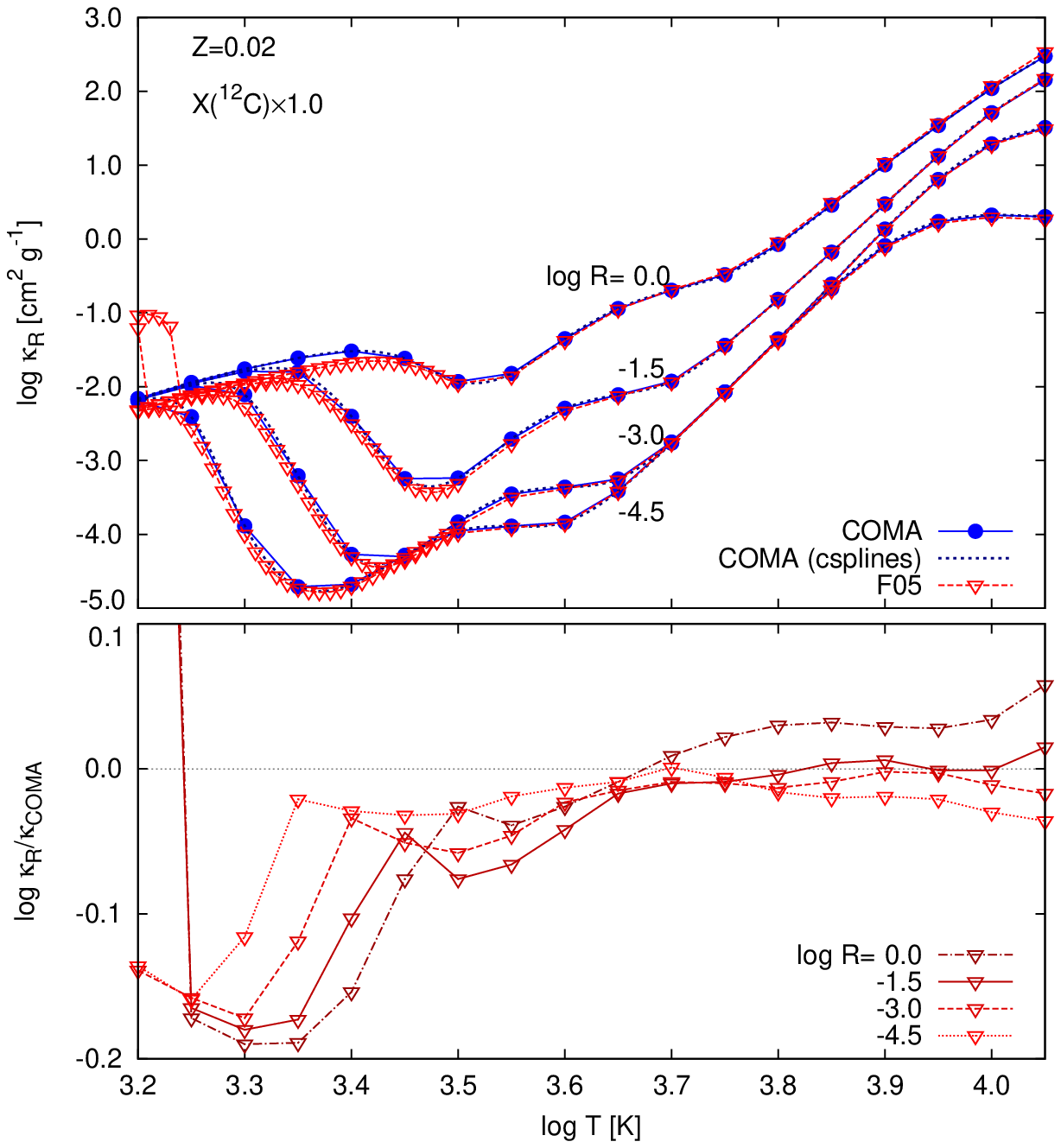}}
   \caption{Comparison of COMA (filled circles) and F05 (empty triangles) values at a metallicity of $Z=0.02$. The top panel shows absolute values, whereas in the bottom panel the differences between F05 and COMA values are indicated on a relative scale. Here and in the following figures, $\kappa_\mathrm{COMA}$ refers to data contained in our database, while the respective comparison values are always labelled $\kappa_\mathrm{R}$. The overall agreement down to $\log T=3.5$ is gratifying. The growing discrepancies towards lower temperatures are most probably due to the deviating set of molecular data in the respective calculations. The steep increase in $\kappa_\mathrm{R}$ in the F05 data below $\log T=3.25$ is due to dust grains that are not accounted for in our data. Below $\log T=3.5$, F05 use a smaller temperature spacing than we do. The regions between the grid points can be reasonably well reconstructed by a cubic spline interpolation in $\log T$ (dotted lines in top panel). In the bottom panel, we compare values at the COMA grid points only.
}
   \label{fig:coma-f05-full-sc}
\end{figure}

Briefly summarised, changes in the chemistry alter the Rosseland mean opacity, where variations in the C/O ratio has the most pronounced effect. An oxygen-rich composition ($\mbox{C/O}<1$) results in a different group of molecules accounting for the opacity than in the carbon-rich case ($\mbox{C/O}>1$). Furthermore, carbon-bearing molecules have different spectral appearances from the oxygen-bearing ones and thus cause $\kappa_\mathrm{R}$ to have a different functional behaviour. \citet{2000A&A...358..651H} provided examples of monochromatic absorption coefficients for carbon-bearing molecules. The only two molecules that contribute significantly for either chemistry are CO and CN.

\begin{figure}
   \centering
   \resizebox{\columnwidth}{!}{\includegraphics{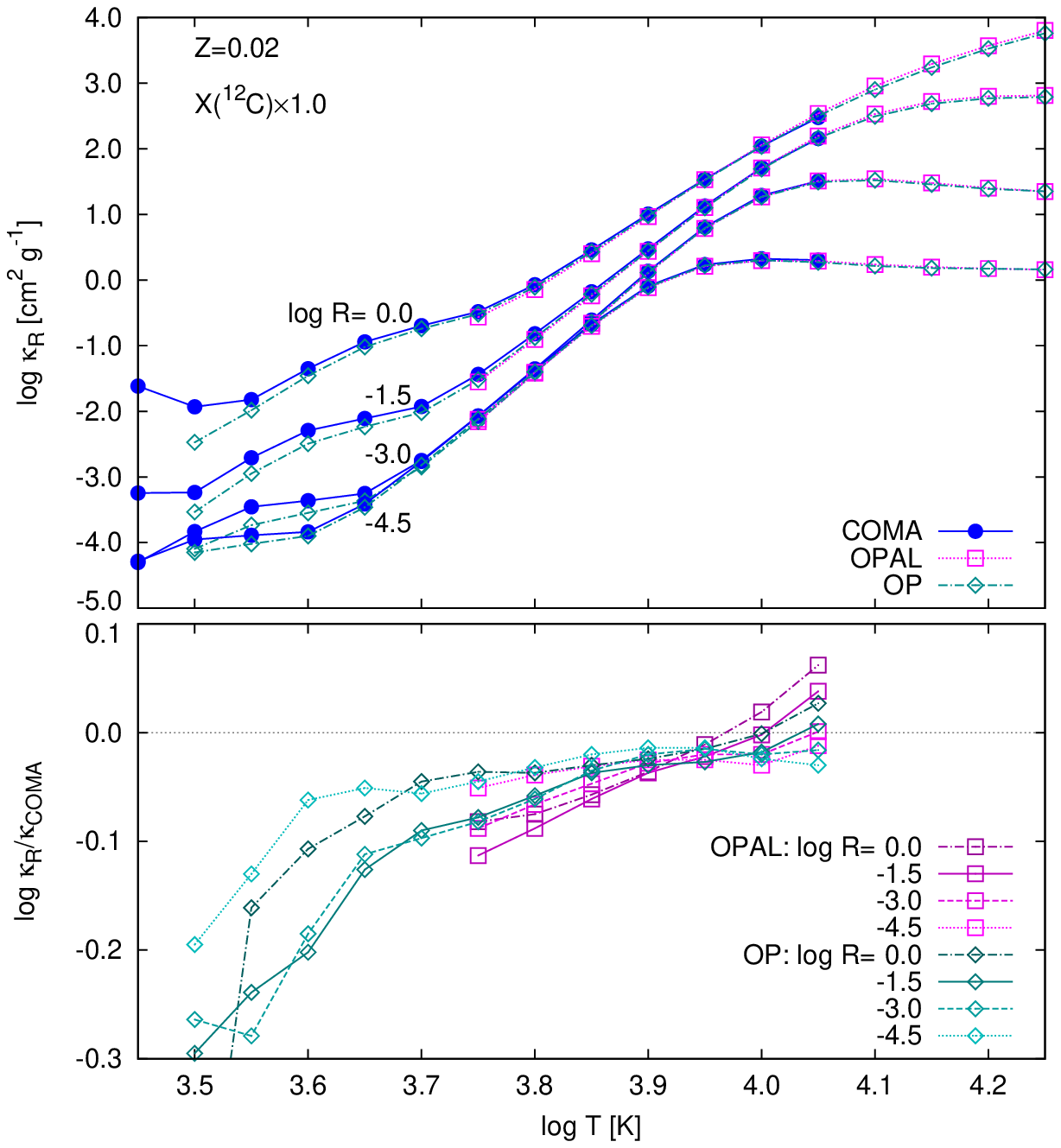}}
   \caption{Comparison of COMA values with OP and OPAL for $Z=0.02$. The top panel shows absolute values, whereas in the bottom panel the differences in the overlapping region are again indicated on a relative scale. The high temperature data do not include the relevant molecular opacity sources. Therefore, the discrepancies increase the lower the temperature becomes. OP data range down to $\log T=3.5$, whereas the OPAL tables end at $\log T=3.75$. The transition between low and high temperature data can be done close to the upper temperature border of the COMA data, where agreement between the values shown here is fairly good. See also Figs.~\ref{fig:coma-op-opal-relative-logR} and \ref{fig:coma-op-relative-logR-Z1E-5}.
   }
   \label{fig:coma-op-opal-full}
\end{figure}

As emphasised earlier, the presented data are primarily relevant to the envelopes of evolved low mass stars. Chemical composition variations due to the TDU acting in AGB stars concern mostly the enrichment in carbon. However, the dredged-up carbon in the envelope can be inputted back into the CN cycle, which partly converts \element[][12]{C} to \element[][14]{N} (see Introduction). By varying the abundance of nitrogen (more precisely \element[][14]{N}), we add a further dimension to our data tables. These alterations have more direct consequences on the behaviour of the Rosseland mean in the sense that an increase in nitrogen always causes an increase in the opacity (in contrast to an increase in carbon, which can lower $\kappa_\mathrm{R}$ for a certain parameter range; see above). Nitrogen is present only within two molecules being considered here, i.\,e. CN and HCN, and can directly influence the Rosseland opacity only via these compounds. Other molecules containing nitrogen indirectly affect the opacity by altering the molecular partial pressures in chemical equilibrium. In the oxygen-rich case, the effect of an increase in \element[][14]{N} is relatively moderate, because the partial pressure of CN is in general low. In the carbon-rich case, the abundance of CN is however much higher, HCN is also present in significant amounts, and, as a consequence, the opacity can increase considerably. These properties of $\kappa_\mathrm{R}$ are illustrated in Fig.~\ref{fig:coma-nitrogen-relative-4} for the two enhancement factors of \element[][14]{N}. On top, we present results for two different metallicities without any carbon enhancement, where the increase in opacity is due to CN only. In the respective bottom panels, the carbon abundance has been enhanced to its maximum value and HCN causes a considerable rise in $\kappa_\mathrm{R}$ at lower temperatures. In the high temperature range, some minor contributions from atomic nitrogen to the opacity are evident.

The results discussed above are condensed into the form of 14 separate files, one for each metallicity.\footnote{The tables are only available in electronic form at the CDS via anonymous ftp to {\tt cdsarc.u-strasbg.fr (130.79.128.5)} or via {\tt http://cdsweb.u-strasbg.fr/cgi-bin/qcat?J/A+A/}. Alternatively, the files will be provided on request.} Each file consists of a header indicating the abundances used, the initial metallicity, the initial mass fractions for \element[][12]{C}, \element[][14]{N}, and the alpha elements, and a look-up table for the true data block. The final file consists of 63 rectangular data arrays, where $\log \kappa_\mathrm{R}$ is tabulated as a function of $\log T$ and $\log R$. The tables are ordered such that the mass fraction $X(\mbox{\element[][12]{C}})$ varies the most rapidly followed by the hydrogen mass fraction and $X(\mbox{\element[][14]{N}})$. For future compatibility, a data field for the alpha element enhancement factor was introduced into the look-up table.

\begin{figure}
   \centering
   \resizebox{\columnwidth}{!}{\includegraphics{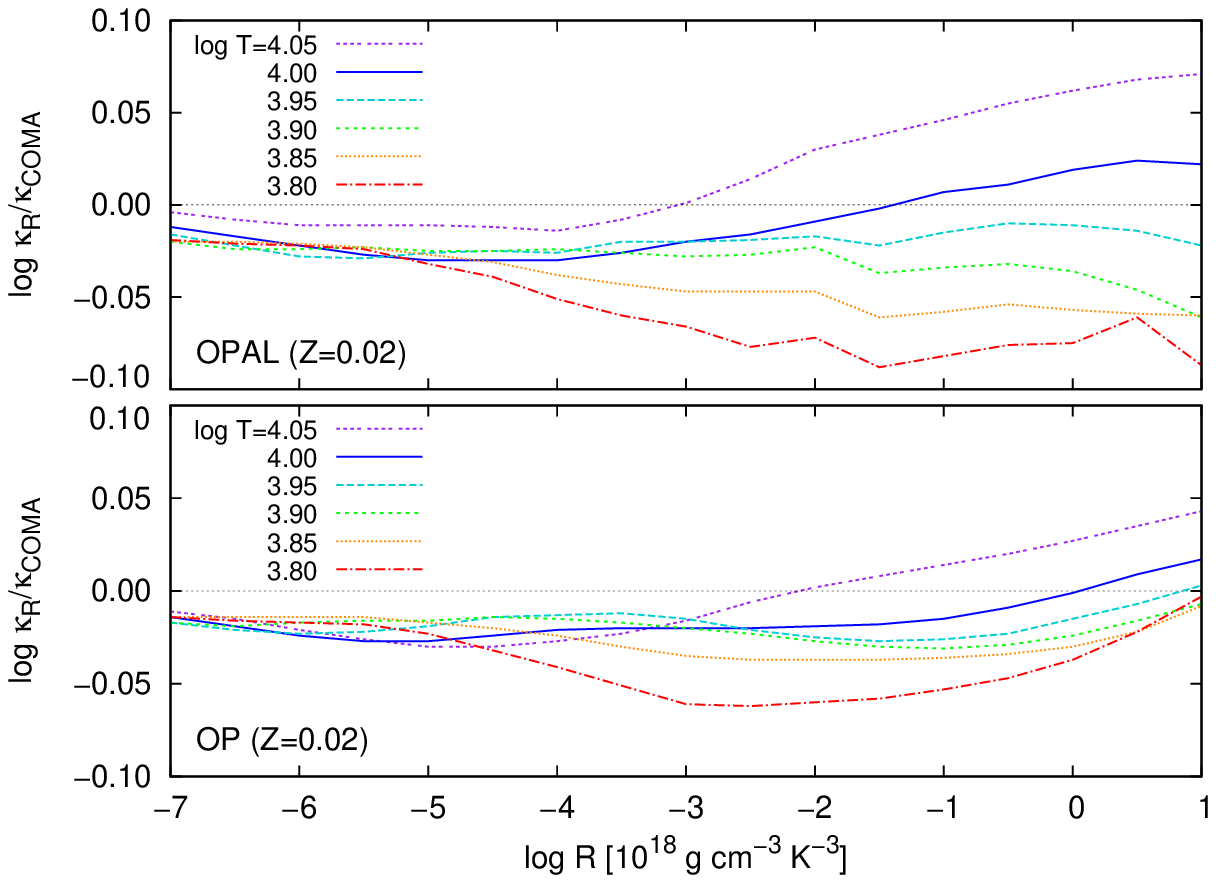}}
   \caption{Comparison of COMA values with OP and OPAL. Our opacity coefficients are systematically higher for a metallicity of $Z=0.02$. This constant offset vanishes for lower metallicities (see Fig.~\ref{fig:coma-op-relative-logR-Z1E-5}), which implies that atomic opacities (lines and continuum) of the metals are the cause. For high values of $\log R$ where the pressure broadening of the spectral lines becomes important, we find increasingly divergent results. The plots appear to imply that the low and high temperature data should be merged around $\log T=4.0$ (solid line).
}
   \label{fig:coma-op-opal-relative-logR}
\end{figure}

\subsection{Comparison with other data}\label{sec:comparison}

We compare our tables based on a scaled solar metal mixture with data from F05 based on the same abundances as in this work. A direct comparison with AF94 is not possible because there are, of course, no tables based on the \citet{2003ApJ...591.1220L} abundances. We refer to F05 for a comparison of AF94 and F05. In the figures, we always depict data from our database as $\kappa_\mathrm{COMA}$, while the respective comparison values are labelled $\kappa_\mathrm{R}$. Despite the numerous differences between the COMA and F05 approach, we find reasonable agreement between both sets of data. For the case shown in Fig.~\ref{fig:coma-f05-full-sc} ($Z=0.02$), the difference between the COMA and F05 values does not exceed 15 per cent for temperatures as low as $\log T=3.5$. The discrepancies at lower temperatures are higher (up to 35 per cent) and can in fact be ascribed to several things. First and foremost, the use of different sets of molecular data in the calculations (cf. our Table~\ref{table:molecules} and their Tables~3 and 4) produces a deviation in the resulting mean opacity coefficients. Second, we adopt a microturbulent velocity of $2.5\,\mathrm{km\,s^{-1}}$, while F05 use $2.0\,\mathrm{km\,s^{-1}}$. The choices for this parameter are (within a certain range that is found for atmospheres of low mass giants) somewhat arbitrary and cause perceptible changes in $\kappa_\mathrm{R}$, especially at lower temperatures. Third, F05 use a denser wavelength grid for the evaluation of $\kappa_\mathrm{R}$. We discuss these issues in more detail in Sect.~\ref{sec:uncertainties}. From a comparison of Fig.~\ref{fig:coma-f05-full-sc} with Figs.~\ref{fig:coma-h2o-c2-relative-logT} (showing a comparable order of magnitude of the deviations) and \ref{fig:coma-xi-f05res-relative-logT}, it is, however, clear that the numerous differences in the physical input data are responsible for the major part of the discrepancies. The resolution and microturbulent velocity influence $\kappa_\mathrm{R}$ not quite as much. The large deviations in the data at the lowest temperatures are due to grain opacity that we do not take into account in our calculations, but dust is usually not formed under equilibrium conditions (as assumed by F05, see Introduction). Moreover, F05 adopted a finer grid in $\log T$ below 3.5. For the oxygen-rich case, a cubic spline interpolation (see Fig.~\ref{fig:coma-f05-full-sc}, dotted lines) on the coarser grid we adopted (and also used by AF94) provides reasonably accurate values.

\begin{figure}
   \centering
   \resizebox{\columnwidth}{!}{\includegraphics{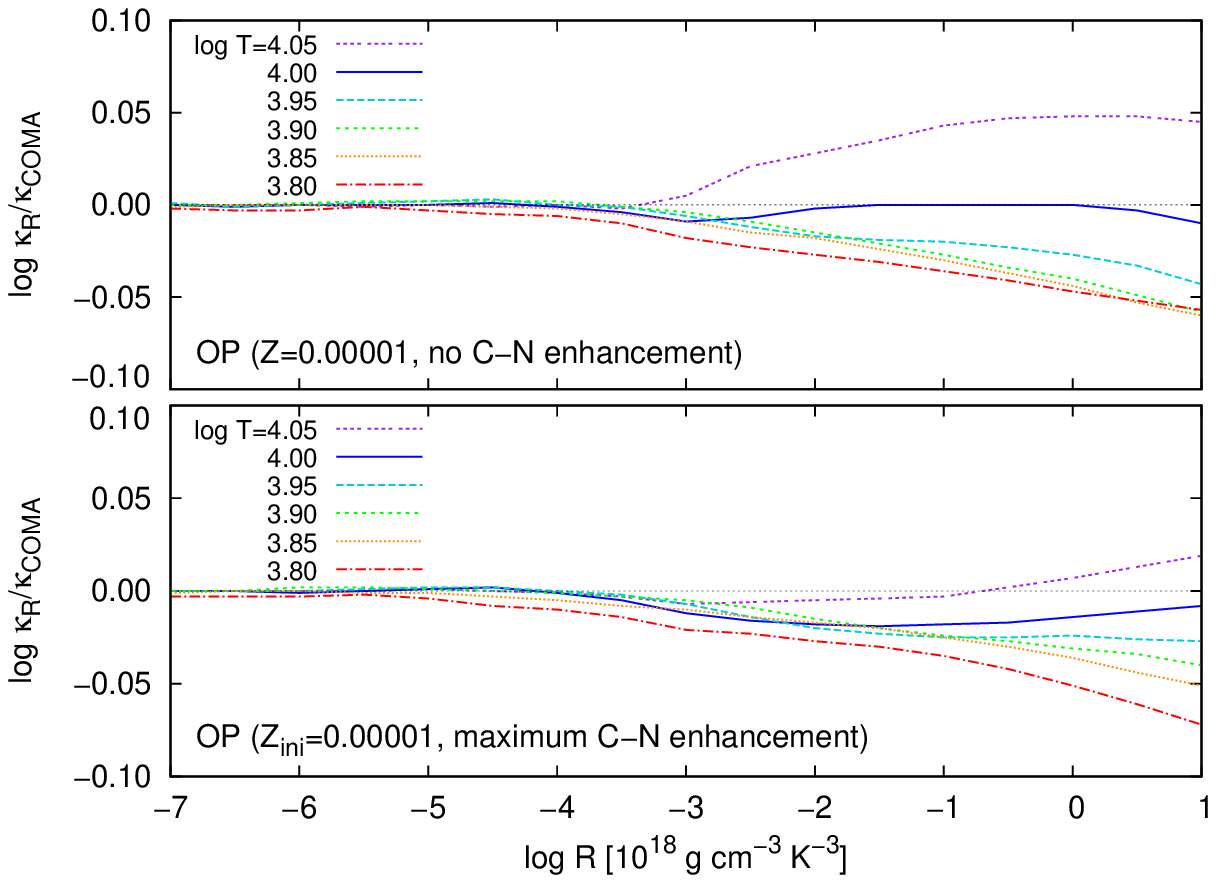}}
   \caption{Comparison of COMA with OP at a low metallicity ($Z=0.00001$) with scaled solar abundances of carbon and nitrogen, and with C and N enhanced to the maximum values (which results in a higher metallicity). The data agree quite well, the discrepancies remain within 5 per cent around $10{,}000\mathrm{\,K}$ (solid line). Again, at high $\log R$ the deviations are higher, obviously due to a deviant treatment of the pressure broadening of spectral lines. From a comparison between OP and OPAL (see text) we draw the same conclusions for OPAL data.}
   \label{fig:coma-op-relative-logR-Z1E-5}
\end{figure}

\begin{figure*}
   \centering
   \resizebox{\textwidth}{!}{\includegraphics{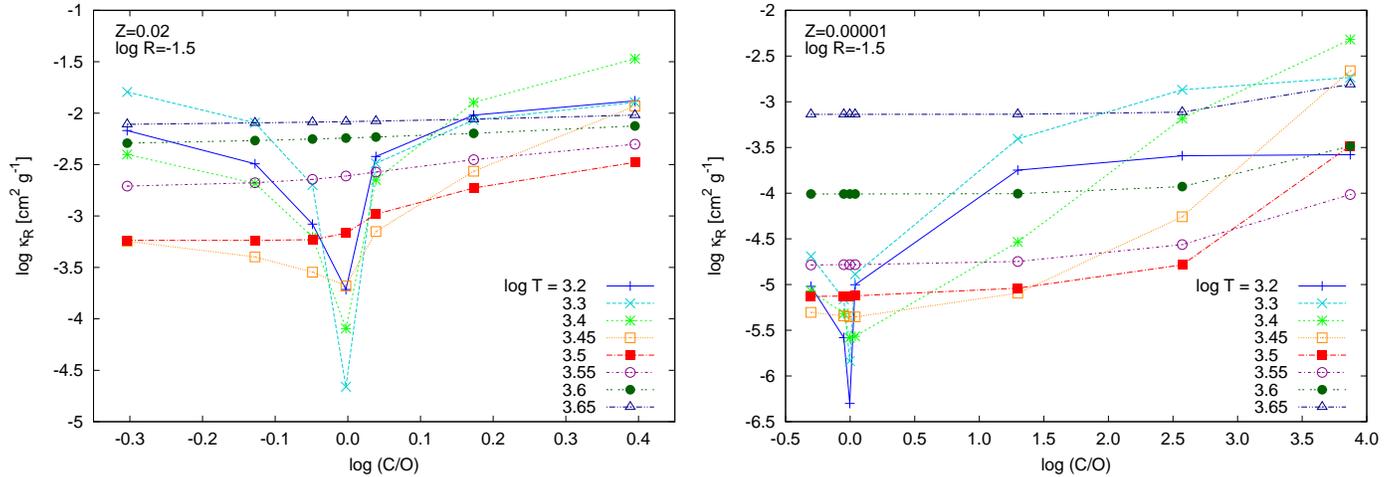}}
   \caption{Evolution of the Rosseland mean as a function of the C/O ratio for different metallicities ($Z=0.02$ and $Z=0.00001$) and a representative value of $\log R=-1.5$. To obtain opacities at enhanced carbon abundances in-between grid points we recommend a linear interpolation scheme in $\log \kappa_\mathrm{R}$ and $\log X(\mbox{\element[][12]{C}})$ as indicated by the lines. The largest relative errors have to be expected at low metallicities and temperatures between $X(\mbox{\element[][12]{C}})\times2.2$ and the successive enhancement factor. For a description of the mechanism that makes $\kappa_\mathrm{R}$ drop to a minimum value at $\mbox{C/O}=1$ and then rise again we refer to Sect.~\ref{sec:results}.
}
   \label{fig:kappa-c-logT-logRm1p5}
\end{figure*}

The comparison with high temperature data such as that from OPAL or OP is limited to the temperature regions where the tables overlap. Moreover, it is this region where a transition between low and high temperature opacities has to be made for applications covering a wide temperature range. OP data stretch down to $\log T=3.5$, whereas the OPAL tables end at $\log T=3.75$. The comparison for a standard scaled solar composition in Fig.~\ref{fig:coma-op-opal-full} shows a growing deviation for lower temperatures because both OPAL and OP do not include molecular absorbers (except H$_2$). This plot indicates that in the region between $\log T=3.8$ and the high temperature end of the COMA data, a smooth transition to high temperature data is possible. Again, from the dimension of the differences, we conclude that these are due to different physical input data rather than other parameters (see Sect.~\ref{sec:uncertainties}). To assess which temperature region renders itself to such a crossover we plot the logarithmic difference between our opacities and those of OPAL and OP, respectively, as a function of $\log R$ in Fig.~\ref{fig:coma-op-opal-relative-logR}. The deviations are in general moderate and the closest agreement is found around temperatures of $10{,}000\,\mbox{K}$. A few comments about the differences are required here. For a metallicity of $Z=0.02$, the values from COMA are systematically higher than both OPAL and OP. This discrepancy almost vanishes at the lowest metallicity considered ($Z=0.00001$, Fig.~\ref{fig:coma-op-relative-logR-Z1E-5}). It must therefore be related to the metallicity, which leaves either the metal lines or the differing continuum opacity as a reason. As for the first possibility, OPAL and OP use a restricted set of the most abundant elements, while we use all data that is contained in VALD. Whether this can cause the differences is a question to be answered by further analysis. Moreover, either OPAL and OP data do not include line broadening due to microturbulence, although this alone cannot account for the difference in $\kappa_\mathrm{R}$.

Beside the constant offset, we recognise a growing discrepancy at increasing $\log R$, which is supposedly to be linked to the pressure broadening of the atomic lines. As mentioned earlier, the damping constants are those from VALD. Additionally, the adopted tabulated hydrogen line profiles could play a role since hydrogen lines contribute significantly to the Rosseland mean at high temperatures (cf. e.\,g. AF94). A deeper investigation of this issue is still to be completed, although the differences remain on an overall satisfactorily small scale.

For a mixture enriched in carbon and nitrogen, we also performed tests. We switched to the lowest metallicity in our database where the effects should be most pronounced due to the high enhancement factors. Here we restrict ourselves to a comparison with OP data since these can be produced quite easily using the software contained in the OPCD (version 3.3). Based on the same mass fractions that we used in COMA, we generated opacity tables with the aforementioned program set and compared them with ours. The results are shown in Fig.~\ref{fig:coma-op-relative-logR-Z1E-5}. We refer to the discussion in the above paragraph concerning the differences but we emphasise again that the quantity $\log \kappa_\mathrm{R}/\kappa_\mathrm{COMA}$ measuring the differences in the logarithmic opacity in dex remains within a reasonable range, i.\,e. $\Delta<0.05\,\mbox{dex}$. In these plots, it is evident that the transition from the COMA coefficients to OP data should occur around a temperature of $\log T=4.0$. From the work of F05 who provided more details about the relation between the low and high temperature opacities, and \citet{2004MNRAS.354..457S} who completed an in-depth comparison between OP and OPAL, we conclude that the above statement also holds for OPAL data.

\subsection{Interpolation}\label{sec:interpolation}

The interface between the opacity tables and the various application codes in which they are used are the interpolation routines. Concerning the temperature and $\log R$, it has become sort of a standard to interpolate in these dimensions using cubic splines \citep[e.\,g. ][]{1993MNRAS.265L..25S,1996MNRAS.279...95S} or quadratic fits (in the Fortran subroutines from Arnold I. Boothroyd\footnote{\texttt{http://www.cita.utoronto.ca/\~{}boothroy/kappa.html}}). The results are quite satisfactory, but we want to emphasise that, in the carbon-rich case, problems can occur. At low values of $\log R$ and a high carbon enhancement, a cubic spline interpolation in the $\log T$ dimension might overshoot and produce spurious results. We strongly advise always checking separately the quality of the fit for each table (or relevant parts thereof) used.
  
The problem now is how to account for the element enhancements. As outlined in Sect.~\ref{sec:results}, the special role of the C/O ratio the parameter range at $\mbox{C/O}=1$ at low temperatures, and $\kappa_\mathrm{R}$ is not a continuous function of the carbon content at this point (Fig.~\ref{fig:kappa-c-logT-logRm1p5}). To resolve the sharp turnaround in the Rosseland mean, we require some grid points close to $\mbox{C/O}=1$. Overall, the number of enhancement factors is too low and the grid too coarse to apply any other interpolation scheme than a linear one. As shown in Fig.~\ref{fig:kappa-c-logT-logRm1p5} for the solar case and at low metallicity, linear interpolation in $\log \kappa_\mathrm{R}$ and $\log X(\mbox{\element[][12]{C}})$ delivers quite gratifying results beyond a certain temperature, where molecules cease to play an important role in determining the value of the Rosseland opacity. The lower the temperature becomes, the sharper the turnaround in the functional behaviour of $\kappa_\mathrm{R}$. Due to the sudden drop in opacity when the amount of carbon and oxygen atoms become approximately equal, linear interpolation misses out a certain fraction of information. At high metallicities the situation is not so serious, although the case shown in the right panel of Fig.~\ref{fig:kappa-c-logT-logRm1p5} ($Z=0.00001$) reveals this shortcoming clearly. Once the element mixture is carbon-rich ($\log \mbox{C/O}>0$), the opacity first increases sharply but flattens at high carbon enhancement values. Due to the construction of our tables (see Sect.~\ref{sec:tabledesign}), the spacing of the enhancement factors in the carbon-rich regime increases at lower metallicities. Between the $X(\mbox{\element[][12]{C}})\times2.2$ and the successive enhancement factor, an additional grid point would be favourable. For the case of nitrogen enhancement, linear interpolation in both $\log \kappa_\mathrm{R}$ and $\log X(\mbox{\element[][14]{N}})$ is a good approximation, because the relation between the nitrogen content and the opacity has a simpler behaviour.

\subsection{Sources of uncertainties}\label{sec:uncertainties}

The definition of the Rosseland mean opacity in Eq.~\ref{eq:rosselandmean} leaves only some ambiguity about how to evaluate this quantity in terms of numerical methods. However, the considerable uncertainties in published opacity coefficients originate in data entering the calculations. In the case of low temperature opacities, there is, in particular, a good amount of physical data of different quality that must be combined into one quantity. The summary in the following paragraphs is not exhaustive but discusses the accuracy of the data presented here and elsewhere.

\begin{figure}
   \centering
   \resizebox{\columnwidth}{!}{\includegraphics{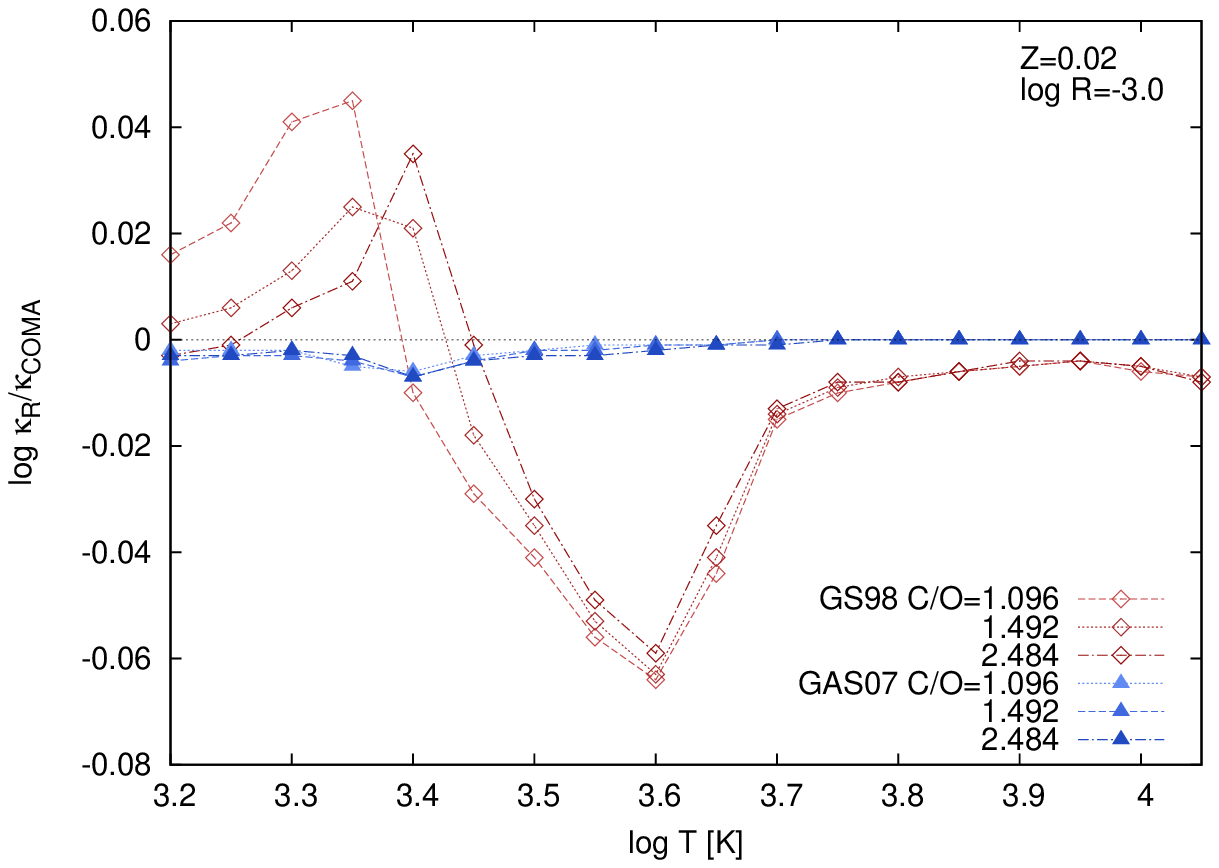}}
   \caption{
    Changes in the Rosseland opacity when using different sets of solar element abundances, for instance \citet[][GS98]{1998SSRv...85..161G} and \citet[][GAS07]{2007SSRv..130..105G}. Data shown here are for the carbon-rich case and $Z=0.02$ at $\log R=-3.0$. The C/O ratios were set to match the values contained in our database using \citet[][L03]{2003ApJ...591.1220L} abundances resulting from $X(\mathrm{\element[][12]{C}})\times2.2,3.0,5.0$. GAS07 is in many respects very similar to L03 (e.\,g. regarding the C, N and O abundances and the share of these elements in $Z$) and thus results in almost identical opacity coefficients. The abundances given by GS98 deviate much more from L03. The differences become manifest in the opacities at high temperatures, and also at low $\log T$ by means of the chemical equilibrium. Other uncertainties discussed in the text have a comparable order of magnitude.
   }
   \label{fig:coma-abundances-relative-logT}
\end{figure}

\begin{figure}
   \centering
   \resizebox{\columnwidth}{!}{\includegraphics{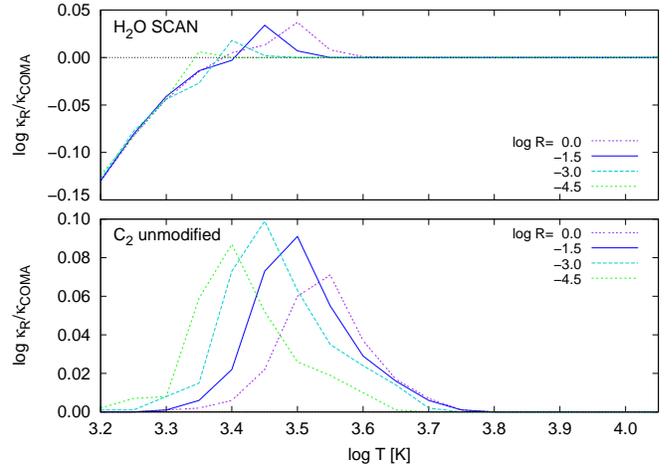}}
   \caption{Uncertainties in the molecular data that affect the mean opacity. Top panel: Using a different line list for water which is the main source of opacity at low temperatures in the oxygen-rich case causes considerable changes in $\kappa_\mathrm{R}$. As a test we substituted the BT2 water line list by the one from SCAN and plot the relative changes compared to the default setup. Bottom panel: For the database described in this paper we scale the C$_2$ line strengths in a certain wavelength region (see text). Not applying this correction results in notable differences in the Rosseland mean opacity of carbon-rich mixtures. ($X=0.7$, $Z=0.02$, no and maximum carbon enhancement in the top and bottom panel, respectively.)}
   \label{fig:coma-h2o-c2-relative-logT}
\end{figure}

\subsubsection{Adopting different solar element abundances}
Hitherto, it has been emphasised that low temperature Rosseland opacities are to a large extent determined firstly by the total metallicity of the element mixture $Z$ and secondly by the C/O ratio. The individual element abundances play a role as well but do not change the opacity on an order-of-magnitude scale when the aforementioned parameters are kept fixed. Concerning the oxygen-rich case, we refer to a discussion of this topic given by \citet{2007ApJ...666..403D}. For the carbon-rich case, we exemplarily calculated opacity coefficients starting from solar element abundances other than \citet[][L03]{2003ApJ...591.1220L}, that is \citet[][GS98]{1998SSRv...85..161G} and \citet[][GAS07]{2007SSRv..130..105G}. The results of our comparison are shown in Fig.~\ref{fig:coma-abundances-relative-logT} (for $Z=0.02$ at $\log R=-3.0$). The solar abundances given by GAS07 are very similar to those of L03, and it is therefore unsurprising to uncover virtually identical opacity coefficients for our test case. The situation is different when we consider the GS98 abundances, which provide higher values for C, N, and O than L03 and GAS07. These elements also make up a higher fraction of $Z$ than in the other cases. In turn, when the metals are scaled to obtain $Z=0.02$, metals apart from C, N, and O, are present in lower amounts than in the L03 case, and thus contribute a smaller fraction to the opacity at high and intermediate temperatures (beyond $\log T=3.5$). At the lowest temperatures, more carbon-bearing molecules, such as C$_3$ and C$_2$H$_2$, are likely to form and produce a higher value of $\kappa_\mathrm{R}$, partially compensating for the lower atomic opacity contribution at intermediate temperatures. The sensitivity of our results to the adopted starting abundances is limited. The size of the differences with respect to the standard case is similar to other uncertainties discussed here. Hence, our data can be used to approximate the Rosseland opacity coefficients for a different set of scaled solar abundances, as long as the CNO abundances do not deviate significantly from the values given in L03.

\subsubsection{Uncertainties in molecular data}
We first consider the molecular line data. For many molecules, there is more than one line list available. The Rosseland mean as a global quantity is insensitive to the precision of line positions so long as the overall opacity distribution is reproduced well. However, there are cases where data from different sources result in altered overall opacity coefficients. As an example, we consider the contribution of water, the major low temperature opacity source in the oxygen-rich case, to $\kappa_\mathrm{R}$. In this work we use the recently published BT2 water line list \citep{2006MNRAS.368.1087B}. An alternative would have been the SCAN database line list from \citet{2001A&A...372..249J}. For the solar metallicity case, we calculated a table utilising this list and illustrate the results in the top panel of Fig.~\ref{fig:coma-h2o-c2-relative-logT}. The discrepancy in the resulting values are as high as 30 per cent. Pronounced differences between the two lists lie in the region where the weighting function in the definition of the Rosseland mean has its maximum, and we ascribe the deviating values of $\kappa_\mathrm{R}$ to this fact.

For the uncertain line data in the carbon-rich case, we mention the modifications to the C$_2$ line data from \citet{1974A&A....31..265Q}. To reproduce carbon star spectra, \citet{2001A&A...371.1065L} proposed a scaling of the $gf$ values in the infrared region \citep[suggested by ][]{Jorgensen1997} based on a comparison with other line lists. More precisely, they scaled the line strengths by a factor of $0.1$ beyond $1.5\,\mathrm{\mu m}$ and left them unchanged below $1.15\,\mathrm{\mu m}$. In-between, they assumed a linear transition. We adopt this method for the calculation of our opacity tables. By not applying this modification to the line strengths, we would have caused an increase of $\kappa_\mathrm{R}$ of roughly 25 per cent at $Z=0.02$ with maximum enhanced carbon (Fig.~\ref{fig:coma-h2o-c2-relative-logT}, bottom panel). The error in these data will have a more significant effect at low metallicities, where one expects a higher enrichment in carbon. From the calculation of mean opacities, we observe a clear need for new and improved C$_2$ line data.

Beside the problems with existing data there are also molecules so far unconsidered that are suspected of providing non-negligible contributions to the opacity. The prime example is C$_2$H, which could be an important opacity source in carbon stars, although to date no line data has existed for this molecule (we refer to \citealp{1995ASPC...78..347G} for an overview).

Another decisive set of input parameters are the chemical equilibrium constants usually depicted by $K_p$. Each constant is in fact a temperature-dependent function setting the partial pressure of a molecule in relation to the product of the partial pressures of the molecule's constituents \citep[cf. e.\,g.][]{1973A&A....23..411T}. \citet{2000A&A...358..651H} pointed out that the literature values for equilibrium constants from different sources could differ strongly at low temperatures. The critical point here is that one has not only to pay attention to the main opacity carriers but also to less abundant molecules competing with them for the same atomic species. \citet{2000A&A...358..651H} referred to TiO and TiO$_2$ as a examples but also reported other molecules for which order-of-magnitude differences in the partial pressures were found using different sets of $K_p$ data. The data that we use is documented in Sect.~\ref{sec:datasources}.

The above examples underline that accurate molecular line data is not only desirable for high resolution applications but also of importance to the calculation of mean opacities. In general, all data used in calculating the Rosseland mean, whether line data or other accompanying data like partition functions or equilibrium constants and continuum sources, must always undergo critical evaluation.

\begin{figure}
   \centering
   \resizebox{\columnwidth}{!}{\includegraphics{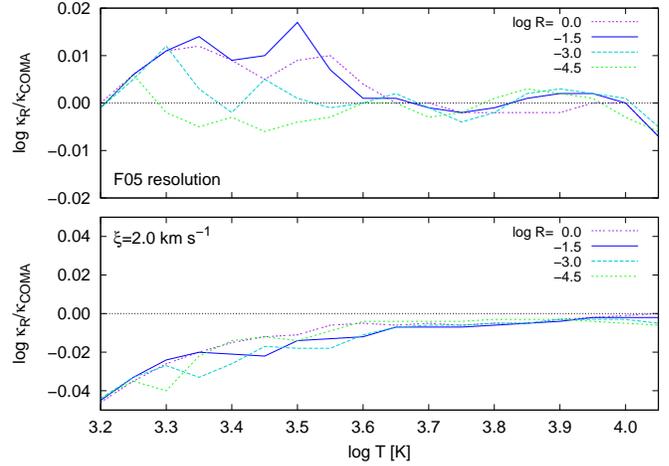}}
   \caption{Influence of variations in resolution or microturbulence on the mean opacity. Top panel: The calculation of $\kappa_\mathrm{R}$ requires to know the monochromatic opacity at a large number of wavelength points. A high spectral resolution assures convergence but induces high computational costs. We use a lower resolution than F05 but the changes in the opacity coefficients are rather moderate especially when compared to other sources of uncertainty. Bottom panel: We adopted a microturbulent velocity of $\xi=2.5\mathrm{\,km\,s^{-1}}$ throughout this work. Changing this value to $\xi=2.0\mathrm{\,km\,s^{-1}}$ like in F05 causes $\kappa_\mathrm{R}$ to drop. ($X=0.7$, $Z=0.02$, no carbon enhancement.)}
   \label{fig:coma-xi-f05res-relative-logT}
\end{figure}

\subsubsection{Numerics and parameters}
Apart from the imprecision due to the physical input data, there are other factors influencing the emerging opacity coefficients. For instance, an error source exists in the wavelength grid for which the opacities are calculated and the integration completed in deriving the opacity mean. Compared to F05, we use a considerably lower spectral resolution. To assess the uncertainties due to this difference, we simulated the resolution of F05, recalculated one of our tables, and compared our results to those for the original case. The differences found were relatively small as shown in Fig.~\ref{fig:coma-xi-f05res-relative-logT} (upper panel). Since the error is low compared to other effects described above, we propose that use of a lower resolution is justifiable because it would reduce considerably the amount of CPU time.

On the other hand, additional physical parameters enter the calculation of $\kappa_\mathrm{R}$, such as the microturbulent velocity $\xi$, which influences the width of the line profiles. The spectral lines are broadened according to the adopted value for $\xi$, which is somewhat arbitrary. Throughout this work, we used a value of $\xi=2.5\mathrm{\,km\,s^{-1}}$ for the generation of our data. Results from previous works on spectra of late-type stars \citep[e.\,g.][]{2002A&A...395..915A,2004A&A...422..289G,2008arXiv0805.3242L} have shown that this is a reasonable assumption. In the work of F05, however, $\xi$ was set to equal $2.0\mathrm{\,km\,s^{-1}}$. Both options are well within the range of values found for AGB star atmospheres (e.\,g. \citealp{1990ApJS...72..387S}). In Fig.~\ref{fig:coma-xi-f05res-relative-logT} (lower panel), we show the results of a test using the F05 value. Since the spectral lines possess a smaller equivalent width at a reduced value of $\xi$, the mean opacity is lower than for the COMA default case.

\subsubsection{Application of the data}
Beyond generating tables, there is further possibilities for introducing uncertainties while applying the data. First, there is the technical problem of interpolating the tabulated values. Compared to previously available data, the situation is worse because there are two more dimensions along which to interpolate, that is the varying amount of carbon and nitrogen. However, on the basis of the above discussion, it is unlikely that too sophisticated interpolation algorithms produce improved accuracy. This is, however, a problem that can in principle be solved by increasing the amount of computer power.

Far more worrying and the largest error source of all is potential misapplication of the data. Strictly speaking, the scope of the tables containing Rosseland mean opacity coefficients are regions where the diffusion approximation for the radiative transfer is fulfilled. In terms of the optical depth, this means $\tau\gg1$ for all wavelengths. One of the main applications of our data will be the outermost parts of an AGB star evolution model. The outer boundary condition is usually set somewhere in the atmosphere ($\log T\leq3.6$), where by definition $\tau\leq1$. In some situations, the Rosseland mean might still be a good approximation for evaluating the radiative energy transport. However, in general it is necessary to use a non-grey radiative transfer method because, due to the molecular absorbers, the spectral energy distribution is strongly wavelength-dependent. We refer to the work of \citet{2003A&A...399..589H}, who demonstrated the shortcomings of a grey treatment of the radiative transfer for dynamical model atmospheres. \citet{2007AIPC..948..195H} investigated the effect of non-grey surface boundary conditions on the evolution of low mass stars and reported noticeable changes to RGB evolution tracks. We thus want to emphasise that our mean opacity tables are meant to provide an interim solution until modelling of non-grey radiative transfer in stellar evolution calculations becomes feasible. 

\section{Conclusions}\label{sec:conclusions}
We have presented a grid of low temperature Rosseland mean opacity tables that take into account variations in the single element abundances of carbon and nitrogen. By gradually enhancing the carbon content of a metal mixture, the molecular contribution to opacity changes significantly due to the altered chemistry. Already within a certain regime (i.\,e. oxygen-rich or carbon-rich), the relative amount of carbon to oxygen has pronounced effects on $\kappa_\mathrm{R}$. More distinctive, however, is the comparison between oxygen-rich and carbon-rich regimes. Different molecules serve as opacity sources in either case and thus result in a qualitatively and quantitatively different Rosseland mean opacity as a function of temperature and density. Changes in the nitrogen abundance also alter the opacity coefficients via certain nitrogen-bearing molecules.

The tables are designed such that an incorporation into existing codes that utilised AF94 or F05 data should be straightforward. Our data cover a wide metallicity range, and the overabundances of carbon and nitrogen are adjusted in each case. We are confident that with these data we provide a tool to simulate the final phases in the evolution of low or intermediate mass stars in more detail. Our data include the effects of the ongoing nucleosynthesis and mixing events in AGB stars in terms of the opacity. As shown in previous papers \citep[e.\,g.][]{2007ApJ...667..489C}, the incorporation of our tables into stellar evolution codes alters the physical properties of the stellar models. Once the star becomes carbon-rich, molecules form that are more opaque than those in an oxygen-rich regime. This in turn results in a steeper temperature gradient. A consequence is, for instance, a decrease in the effective temperature in stellar evolution models. The stellar radius increases, and the average mass-loss rate increases and erodes the envelope mass at a faster rate. \citet{2003PASA...20..389S} showed that a change in the envelope mass (as well as a change in the core mass) affect fundamental properties of AGB stars, e.\,g. the strength of the thermal pulses and the total amount of the mass dredged up. It will also be interesting to see how different mass-loss prescriptions interact with the newly calculated opacity coefficients, since these issues are physically closely coupled.

In the future, we plan to extend our tables to contain data about the enrichment in alpha elements. We must emphasise, however, that the data provided in the course of the current and future work must be seen as a transitional solution to the treatment of molecular opacity in AGB star envelopes and atmospheres. Due to the band structure of molecular absorption, mean opacities will yield inaccurate results. The past results from static and dynamical model atmosphere calculations demonstrate the importance of a frequency-dependent radiative transfer. Our data promise to bridge the gap until these methods are employed in stellar evolution models.

Finally, we emphasise that for forthcoming extensions of this database, it would be desirable to obtain extensive response from the community. Comments and criticisms that can lead to an improvement in the quality of the data are highly welcome. 

\begin{acknowledgements}
MTL and BA acknowledge funding by the Austrian Research Fund FWF (projects {P-18171} and {P-19503}). BA received financial support from the University of Padova (Progetto di Ricerca di Ateneo {CPDA052212}). MTL has been supported by the Austrian Academy of Sciences (DOC programme). We also want to thank Sergio Cristallo who pointed out the need for the data presented in this work and provided us with many useful comments. Christian St\"utz is thanked for delivering an updated and ready-to-use set of VALD data.  
\end{acknowledgements}

\end{document}